\documentclass[twocolumn,amsmath,amssymb,floatfix,pra,12pts,graphicx]{revtex4-2}
\usepackage[caption=false]{subfig}  

\usepackage{overpic} 
\usepackage[utf8]{inputenc}
\usepackage{epstopdf}
\usepackage{dcolumn}
\usepackage{bm}
\usepackage[mathlines]{lineno}
\usepackage{xcolor}
\usepackage{soul}
\usepackage{bm}
\usepackage{bbm}
\usepackage{amsmath,amsfonts,amssymb}
\usepackage{extarrows} 
\usepackage{mathtools}
\usepackage{enumitem} 
\usepackage{xr}
\usepackage{xcolor}

\usepackage{hyperref}
\hypersetup{
colorlinks,citecolor=red,
filecolor=blue,
linkcolor=blue,
urlcolor=black
}

\definecolor{sph}{rgb}{0.0588, 0.3216, 0.7294} 
\definecolor{ppk}{rgb}{1.0, 0.4549, 0.0902} 
\newcommand{\nc}{\newcommand}
\nc{\nn}{\nonumber}
\nc{\txt}{\textrm}
\newcommand{\beq}{ \begin{equation} }
\newcommand{\eeq}{ \end{equation} }
\nc{\txtsup}{\textsuperscript}
\nc{\txtsub}{\textsubscript}
\nc{\calL}{\mathcal{L}}
\nc{\U}{\mathcal{U}}
\nc{\T}{\mathcal{T}}
\nc{\E}{\mathcal{E}}
\nc{\cH}{\mathcal{H}}
\nc{\sect}[1]{{\sl #1 .--}}

\newcommand\nya[1]{\{ #1 \}}

\nc{\RV}[1]{\textcolor{blue}{#1}}
\nc{\YD}[1]{\textcolor{red}{#1}}

\nc{\cC}{\mathcal{C}}
\nc{\cI}{\mathcal{I}}

\begin{document}
\title{
Qlustering: Harnessing Network-Based Quantum Transport for Data Clustering 
}

\author{Shmuel Lorber and Yonatan Dubi}
\affiliation{Department of Chemistry, Ben Gurion University of the Negev, 1 Ben-Gurion Ave, Beer Sheva 8410501, Israel }
\date{\today}
\email{shmuell@bgu.ac.il; jdubi@bgu.ac.il}
\begin{abstract}
We introduce Qlustering, a quantum-inspired algorithm for unsupervised learning that leverages network-based quantum transport to perform data clustering. In contrast to traditional distance-based methods, Qlustering treats the steady-state dynamics of quantum particles propagating through a network as a computational resource. Data are encoded as input states in a tight-binding Hamiltonian framework governed by the Lindblad master equation, and cluster assignments emerge from steady-state output currents at terminal nodes. The algorithm iteratively optimizes the network’s Hamiltonian to minimize a physically motivated cost function, achieving convergence through stochastic updates. We benchmark Qlustering on synthetic datasets, a localization problem, and real-world chemical and biological data, namely subsets of the QM9 molecular database and the Iris dataset. Across these diverse tasks, Qlustering demonstrates competitive or superior performance compared with classical methods such as k-means, particularly for non-convex or high-dimensional data. Its intrinsic robustness, low computational complexity, and compatibility with photonic implementations suggest a promising route toward physically realizable, quantum-native clustering architectures.
\end{abstract}
   
\maketitle

\section{INTRODUCTION}
The search for efficient, non-Turing computing architectures that harness quantum mechanics for machine learning (ML) has gained renewed momentum in recent years. This resurgence is driven by breakthroughs such as quantum kernel estimation for classification tasks and quantum-enhanced support vector machines \cite{havlivcek2019supervised}, alongside the development of quantum generative adversarial networks (QGANs) \cite{lloyd2018quantum,dallaire2018quantum}. These efforts pursue two intertwined goals: to reduce the computational burden of classical ML by leveraging the structure of physical systems, and to overcome the limitations imposed by noise in current quantum hardware. While much attention has been paid to the algorithmic front, hardware innovation is increasingly recognized as essential to achieving robust quantum ML performance. Recent developments in silicon photonics and integrated photonic chips have enabled unprecedented scalability and precision in quantum hardware. One notable example is programmable nanophotonic processors (PNPs) \cite{harris2018linear}, which offer new opportunities for implementing quantum ML systems in noise-resilient architectures.\\

A particularly promising yet underexplored direction in quantum machine learning involves harnessing quantum transport phenomena in open quantum systems to carry out information processing tasks. Rather than modeling physical devices, this approach treats the dynamics of quantum systems as computational resources, a concept now known as "computing with physical systems" \cite{stern2023learning, mohseni2022ising, mcmahon2023physics, wetzstein2020inference, PhysRevX.11.021045} . For example, Dalla Pozza et al. \cite{dalla2020quantum} demonstrated that such systems can be configured for quantum state discrimination, reaching the Helstrom bound - the theoretical optimum for this task. Building on this idea, Wang et al. \cite{wang2022implementation} proposed using quantum networks for classification problems. In our recent work \cite{lorber2024using}, we extended this paradigm by showing that current-based quantum neural networks can outperform classical models across a variety of classification tasks, including mathematical, physical, and real-world chemical datasets.\\

In this work, we extend the work introduced \cite{lorber2024using}, and describe a quantum algorithm for unsupervised learning, specifically clustering. Built upon transport of quantum particles through a network. We refer to this method as "Qlustering". As detailed in Section~\ref{setup}, the algorithm leverages the interplay between the network’s structure, quantum particle propagation, and the resulting output currents at terminal nodes.

Clustering algorithms aim to identify natural groupings within a dataset based on inherent similarities or patterns, without prior knowledge of class labels \cite{shalev2014understanding}. Qlustering operates as a distance-based method that groups inputs based on their relative positions in Hilbert space. It is designed to partition any dataset - whether quantum or classical - whose elements differ spatially in an $N$-dimensional space once embedded in an $N$-dimensional Hilbert space. However, as demonstrated in Section~\ref{RESULTS}, Qlustering outperforms traditional distance-based techniques and functions as a genuinely quantum-native algorithm.

We illustrate its capabilities through two case studies involving structurally distinct, non-distance-based datasets, showing Qlustering's flexibility and generality. A third case study, presented in Section~\ref{QM9}, applies Qlustering to real-world chemical data from the QM9 dataset. Finally, in Section~\ref{summary}, we summarize by compare Qlustering to classical clustering algorithms and discuss its advantages as well as its current limitations.

\section{SETUP AND FORMULATION}
\label{setup}
\begin{figure} 
\centering
\includegraphics[width=\linewidth, trim={0 130 0 6cm}, clip]{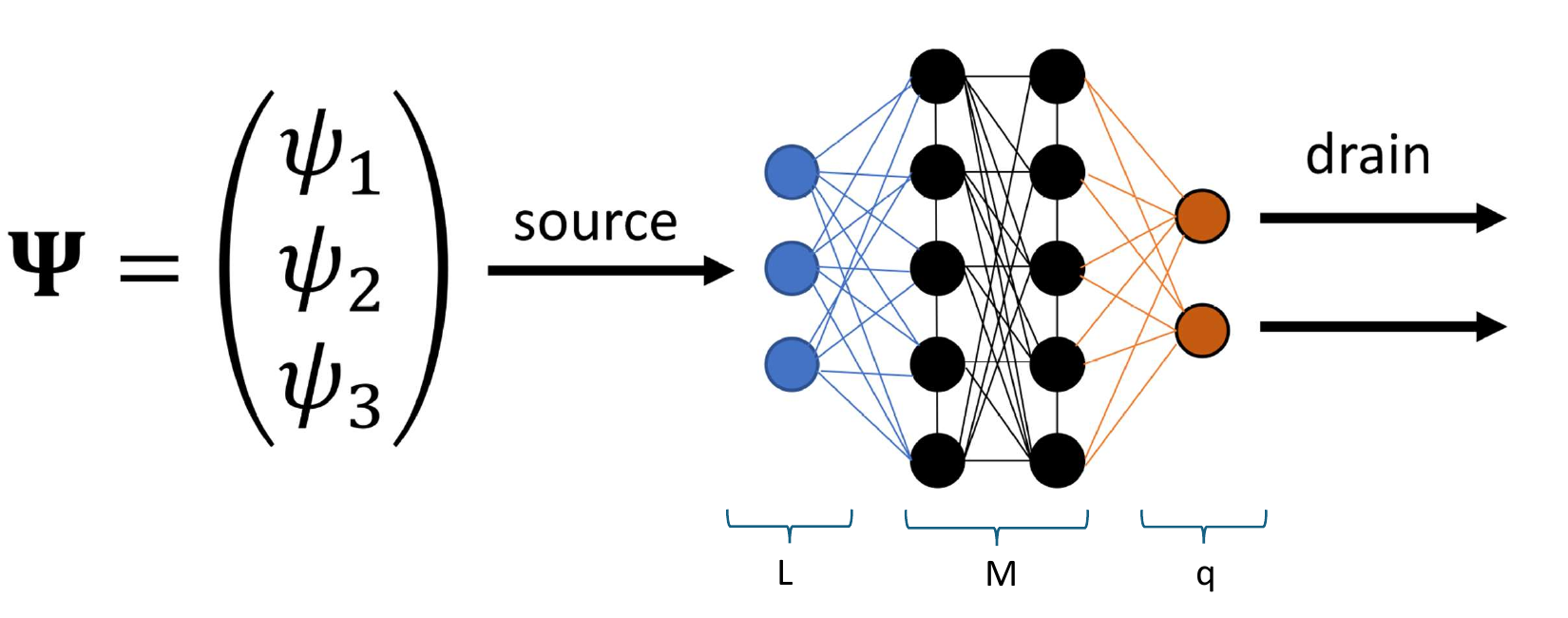}
\vspace{-0.5em} 
\caption{\textbf{Schematic representation of the Qlustering network.}
An input state vector $\Psi$ of dimension $L$ is injected into the network through the input nodes. In this figure, $L=3$. The propagation, injection, and extraction of a particle are modeled by Eq.~\ref{eqn1}. Here, $L$ denotes the number of input nodes and also the dimensionality of the state vectors, $q$ is the number of output nodes (i.e., clusters), and $M$ is the number of hidden nodes. After reaching a steady state, the current from each output node is computed. The state is then assigned to the group corresponding to the output node with the highest current.}
\label{figure 2}
\end{figure}

\normalsize \textbf{The Qlustering algorithm.--} 
As pointed above, the Qlustering algorithm is a "physical computing", i.e. the natural operation of a physical device.   We begin by describing the physical and structural properties of the Qlustering device. Given a set of $N$ state vectors, $\left\{ \Psi_n \right\}$, in an $L$-dimensional Hilbert space, and a predetermined number of clusters $q$, the task is to assign each $\Psi_n$ to its corresponding class $\mathcal{C}[\Psi_n]$. The network is constructed as follows (see Fig.~\ref{figure 2}): consider a system-environment combination in which the system is the network, and the environment acts as a reservoir. The $L$ input nodes and $q$ output nodes are connected to the environment (such that particles can flow into the input nodes and out of the output nodes) but are not directly linked to each other. Between the input and output layers, there are $M$ "hidden" nodes. The dynamics of particles in the network are governed by a tight-binding Hamiltonian of the form $\cH=\sum_{i,j} h_{ij} c^\dagger_i c_j + \text{h.c.}$, where $c^\dagger_i$ ($c_i$) creates (annihilates) a particle at node $i$. A full physical and mathematical description is given in the next section; for now, it is essential to note that the Qlustering algorithm is parameterized by the hopping terms \( h_{ij} \) of \( \mathcal{H} \).

To Qluster a given data element $\Psi_n$, one excites the input nodes (in a manner that encodes $\Psi_n$) and allows the particle to propagate freely within the network until it reaches a steady state. Once steady state is achieved, the output current vector is computed as $J[\Psi_n]=\begin{bmatrix} j_1 \\ j_2 \\ \dots \\ j_q \end{bmatrix}$. The output node $i \in \{1, \dots, q\}$ that yields the highest current $j_{\text{max}}$ determines the class to which $\Psi_n$ is assigned. After processing the entire dataset, clusters are defined by grouping together the state vectors associated with the same output node corresponding to $j_{\text{max}}$. 
The clustering behavior, dictated by the hopping terms $h_{ij}$ of the Hamiltonian $\cH$, is optimized through an iterative procedure described below.\\

The Qlustering algorithm proceeds as follows:
\begin{enumerate}
    \item Initialize the Hamiltonian $\cH$ randomly.
    \item Propagate each state vector $\Psi_n$ through the network and calculate its steady-state current $J[\Psi_n]$ (a vector of length $q$). Importantly, we identify for each $\Psi_n$ the output node with maximal current, $i_{mx}(n)$.
    \item Once all currents are computed, i.e., $J=J[\Psi_1],J[\Psi_2], \dots, J[\Psi_N]$, evaluate the clustering quality using the following cost function (CF):
    \begin{equation}
    \label{CF}
    CF(J,N)=\sum^N_{n=1} \left(I_n - J_n \right)^2~.
    \end{equation}
   Here, $J_n=J[\Psi_n]$, and $I_n$ is a vector of length $q$ (same length as $J_n$) defined as $I_n(i)=\delta_{i,i_{mx}(n)}$ (where $\delta_{i,j}$ is the Kronecker delta), i.e. it is a vector 
   with all entries zero except a single one at the index where corresponding to the index of the maximal current output for $\Psi_n$ and the index $n$ runs over the full dataset
    
    \item Randomly select an entry in $\cH$ and modify it to produce a new Hamiltonian $\cH^{(2)}$. 
    
    \item Repeat Steps~2--3. When the Hamiltonian changes, the currents $J_n$ are modified, and consequently the vectors $I_n(i) = \delta_{i,i_{mx}(n)}$ and the cost function also change. This contrasts with the classification cost function presented in~\cite{lorber2024using}, which remained fixed throughout the training process.
    
    \item If the value of the new cost $CF^{(2)}$ is lower than the previous cost $CF^{(1)}$, retain $\cH^{(2)}$; otherwise, revert to $\cH^{(1)}$.
    \item Repeat steps 4–6 until convergence is achieved.
\end{enumerate}

To accelerate convergence, we employ multiple particles in Step~4—that is, we draw several values for the selected entry in $\cH$ and compute, in parallel, the corresponding currents and cost function ($CF$) for each.

\begin{figure}[t]
    \centering
    \captionsetup[subfloat]{labelformat=empty} 

    \begin{minipage}[t]{0.3\linewidth}
        \centering
        \textbf{(a)} \\[0.3em]
        \includegraphics[width=\linewidth]{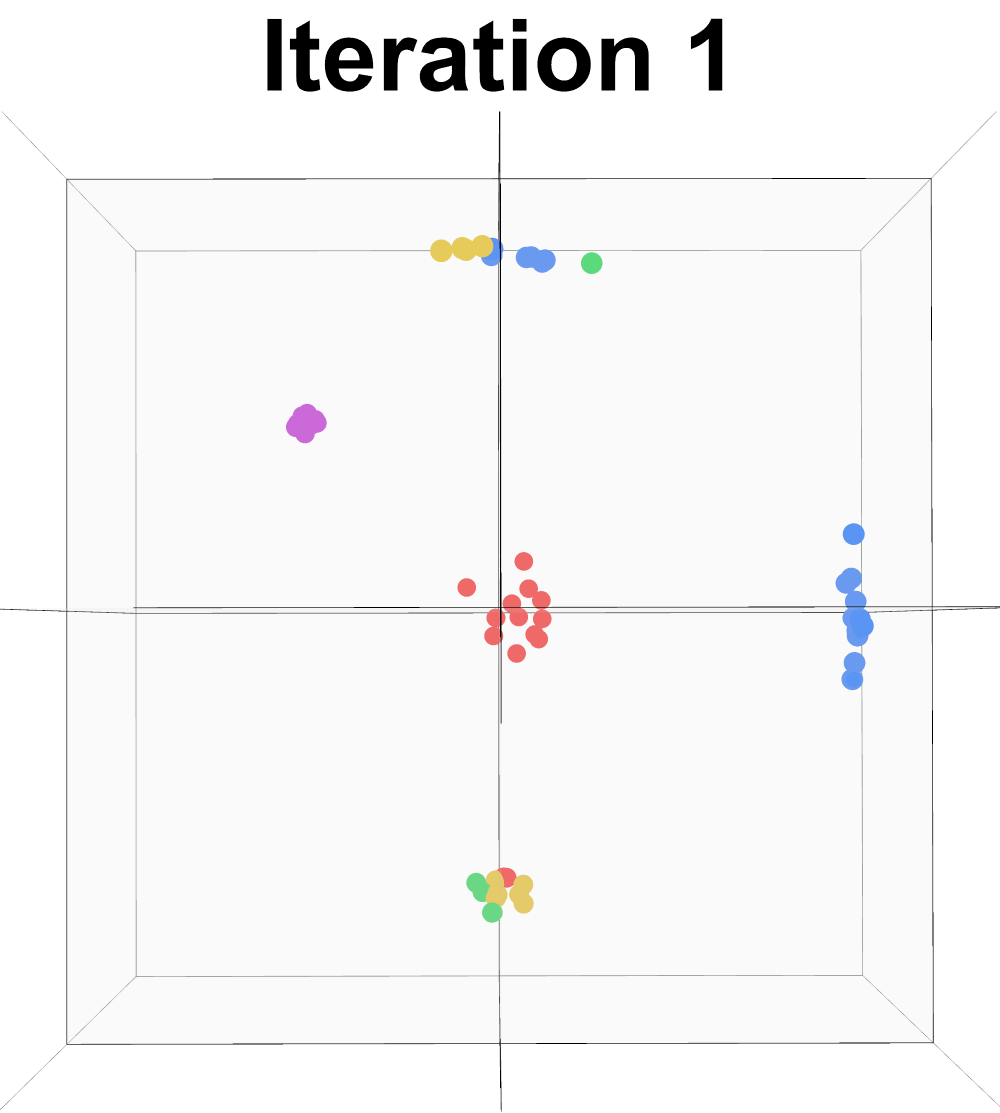}\\[0.5em]
        \includegraphics[width=\linewidth, trim={0 250 0 0cm}, clip]{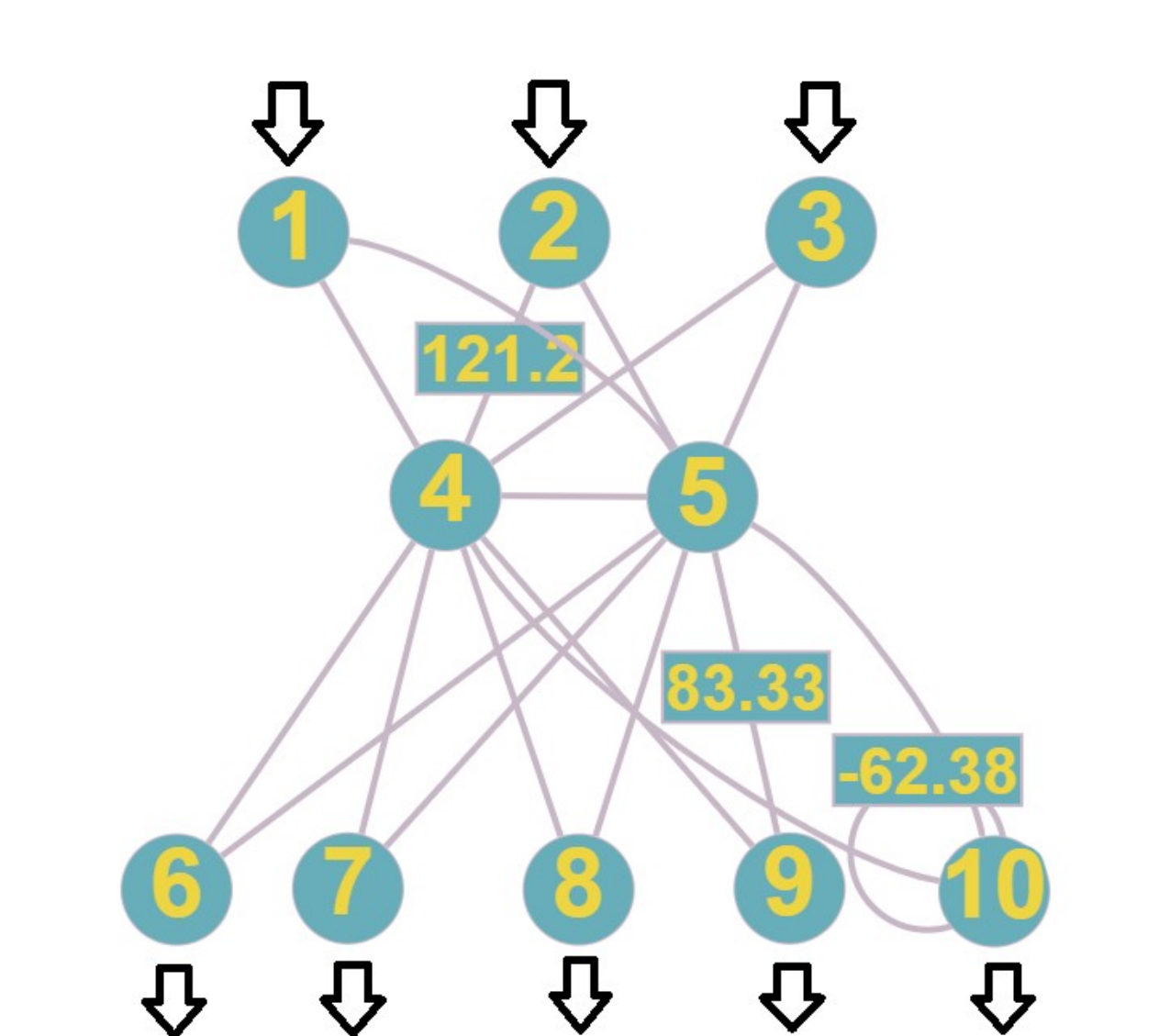}
        \label{fig:col-a}
    \end{minipage}
    \hfill
    \begin{minipage}[t]{0.3\linewidth}
        \centering
        \textbf{(b)} \\[0.3em]
        \includegraphics[width=\linewidth]{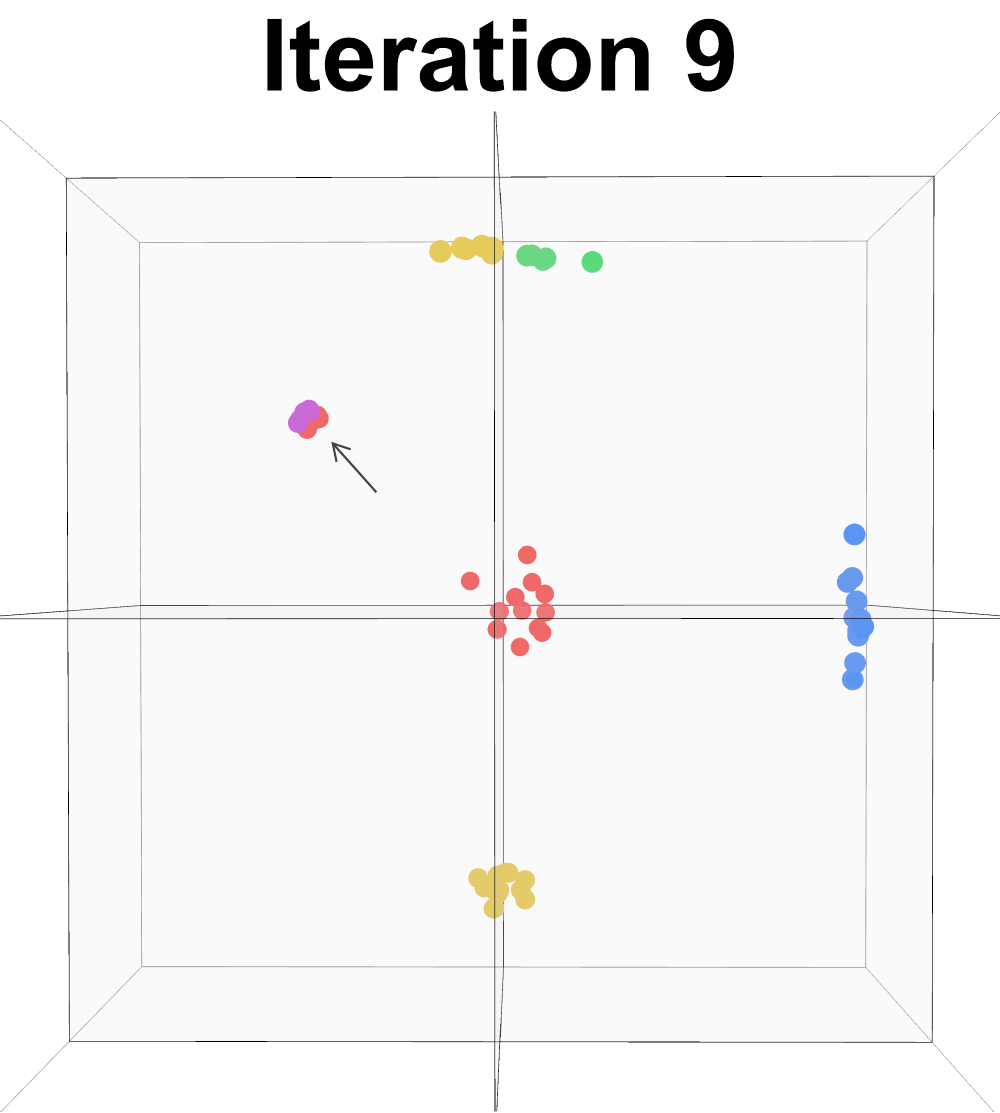}\\[0.5em]
        \includegraphics[width=\linewidth, trim={0 246 0 0cm}, clip]{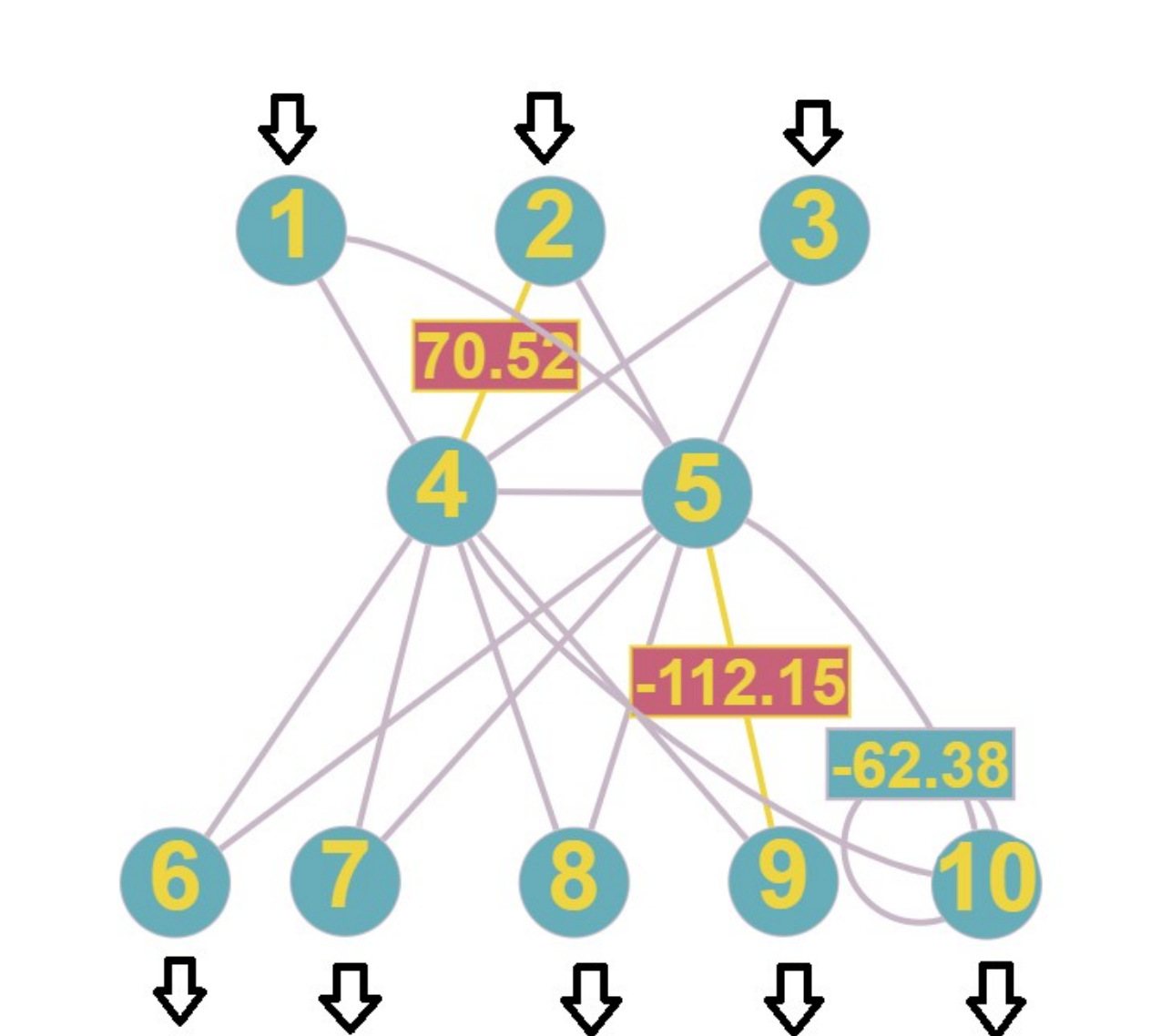}
        \label{fig:col-b}
    \end{minipage}
    \hfill
    \begin{minipage}[t]{0.3\linewidth}
        \centering
        \textbf{(c)} \\[0.3em]
        \includegraphics[width=\linewidth]{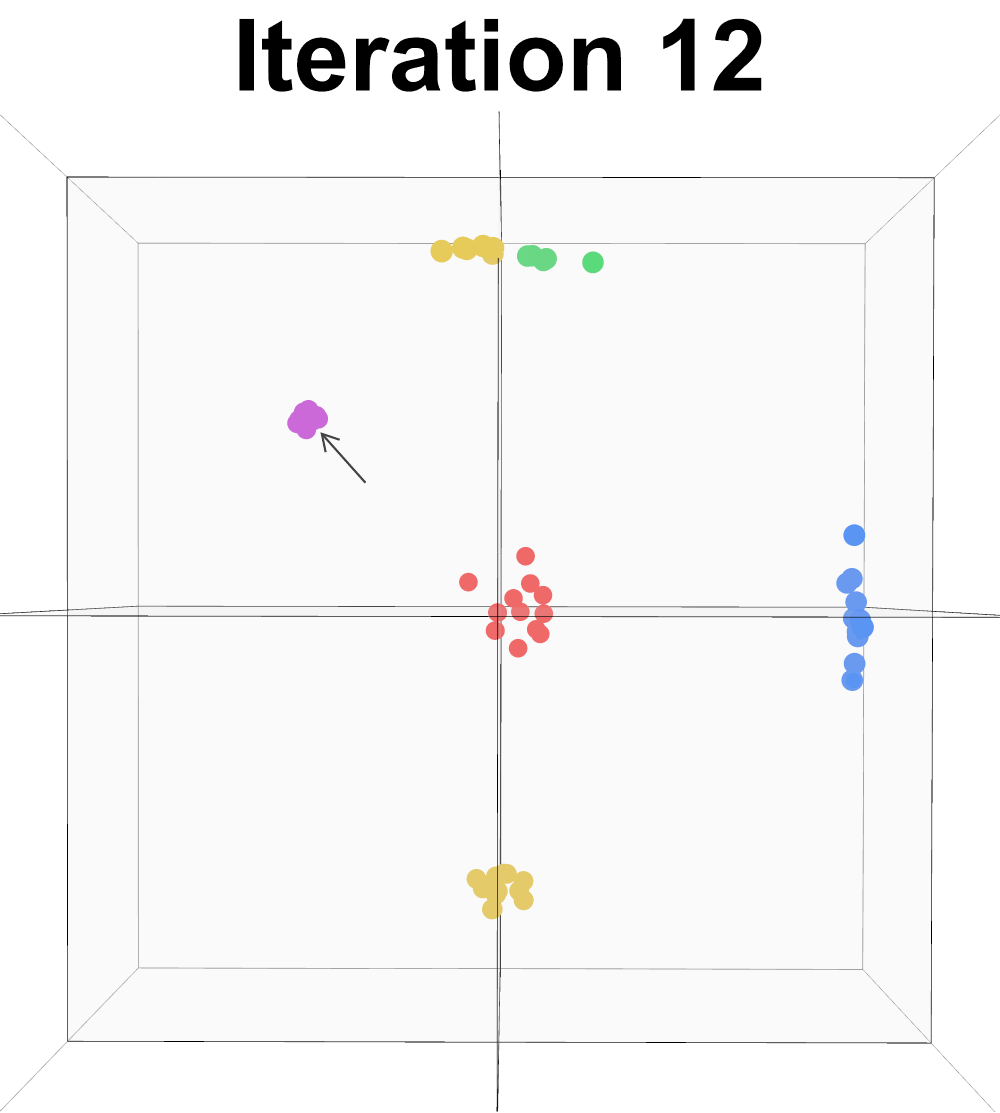}\\[0.5em]
        \includegraphics[width=\linewidth, trim={0 246 0 0cm}, clip]{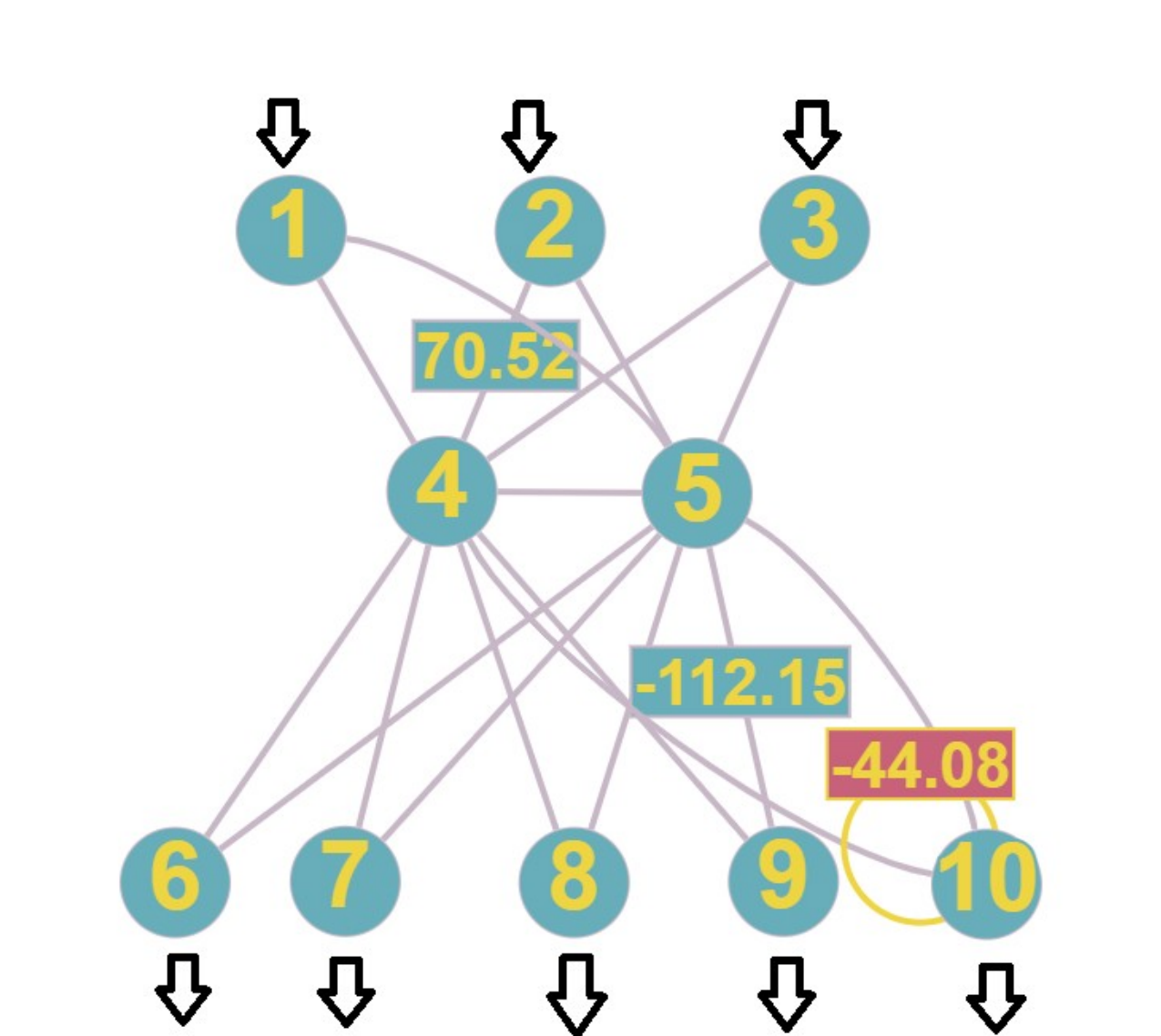}
        \label{fig:col-c}
    \end{minipage}
    
\caption{Frames captured from the Qlustering process applied to 60 state vectors in a 3-dimensional Hilbert space, grouped into 5 distinct clusters. Upper panels illustrate the spatial distribution of the vectors in the space, while the bottom panels illustrate the Qlustering network with its 3-2-5 node structure. The fitting process runs over iterations $t = 1, \dots, T$, where $T$ is a predefined number of steps. Panel (a) shows the initial state at $t = 1$, where 3 clusters are already correctly grouped. The corresponding Hamiltonian for this initial step is displayed in bottom panel of (a). In each subsequent iteration, a random entry in the Hamiltonian $\mathcal{H}$ is selected and modified. If the new Hamiltonian yields a lower cost function $CF$ (Eq.~\ref{CF}), it is retained and used in the next iteration. Bottom panel (b) presents the Hamiltonian at $t = 9$, with the two changes from the initial Hamiltonian highlighted in yellow. The resulting clustering at $t = 9$ is shown in upper panel (b). The panels of (c) illustrate the clustering and Hamiltonian at iteration $t = 12$.
}
    \label{figure 1}
\end{figure}

To illustrate the method (see Fig.~\ref{figure 1}), consider a set of states in an $L=3$ Hilbert space distributed into $q=5$ clusters. The algorithm is provided with the number of clusters $q$ and the state vectors $\Psi = \Psi_1, \Psi_2, \dots, \Psi_N$. An initial random Hamiltonian - depicted graphically in Fig.~\ref{figure 1}(f) - is used to simulate the network and compute the initial steady-state currents. This defines the first clustering, as shown in Fig.~\ref{figure 1}(a), where each point represents a state vector’s position in Hilbert space and its color denotes the assigned cluster. In the next iteration, a single element of $\cH$ is modified; the updated clustering is shown in Fig.~\ref{figure 1}(b). Fig.~\ref{figure 1}(g) displays the Hamiltonian after nine iterations, differing in two entries from the initial one. The process continues until convergence or a predefined number of iterations is reached. The final clustering is shown in Fig.~\ref{figure 1}(e). A full video of the Qlustering process is available in the Supplementary Information. 

\onecolumngrid

\twocolumngrid

{\bf The quantum network.--} As mentioned, we consider a one-particle tight-binding Hamiltonian of the form $\mathcal{H} = \sum_{i,j} h_{ij} c^\dagger_i c_j + \text{h.c.}$ This general framework can model a variety of quantum transport networks, including (but not limited to) electron transport in quantum dots \cite{sarkar2022emergence}, exciton transport in biological systems \cite{zerah2021photosynthetic}, photon propagation in waveguides \cite{mookherjea2001optical,chen2021tight}. and programmable nanophotonic processors \cite{harris2017quantum} \\

To model propagation through the network, we use the Gorini-Kossakowski-Sudarshan-Lindblad (GKSL) quantum master equation \cite{lindblad1976generators,gorini1976completely}:
\begin{eqnarray}
\dot \rho &=& -i[H, \rho] + \sum_k \left( V_k^{\dagger} \rho V_k - \frac{1}{2}(V_k^{\dagger} V_k \rho + \rho V_k^{\dagger} V_k) \right) \\ \nonumber
&=& -i[H, \rho] + \mathcal{L}[\rho]
\label{eqn1}
\end{eqnarray}
Here, $\rho$ denotes the system's density matrix, and $V_k$ are the Lindblad operators. In this setup, we define two Lindblad operators. The input operator is given by $V_{\text{in}} = \gamma_{\text{in}}^{1/2} \sum_{i=1}^{L} \Psi(i) c^\dagger_i$ \cite{zerah2021photosynthetic}, where $c^\dagger$ is the creation operator, $\gamma_{\text{in}}$ is the input dissipation rate, and $\Psi(i)$ is the $i$-th component of the input state vector $\Psi$. This operator represents a superposition of $\Psi$ injected into the input nodes.

The output operator is defined as $V_{\text{out},n} = \gamma_{\text{out}}^{1/2} c_r$, where $\gamma_{\text{out}}$ is the output dissipation rate and $c_r$ is the annihilation operator at the output node $r$. See Fig.~\ref{figure 2} for a schematic illustration.

The steady-state current resulting from this setup is computed as detailed in Section~\ref{METHODS}. This current serves as the input to the cost function defined in Eq.~\ref{CF}. After tuning the network parameters and clustering the dataset, the next step involves evaluating the structure of the data and validating the algorithm’s performance.

\textbf{Data structure and validation.--} To assess the performance of the Qlustering algorithm, it is first necessary to define the nature and structure of the data being clustered. In this study, following Fahad et al.~\cite{fahad2014survey}, we used three types of data: synthetic data with tunable convexity, physical data with relatively high dimensionality (10 parameters per data point), and two examples of real-world datasets - QM9 and Iris.\\

To evaluate clustering performance, we employ external validation metrics such as the Rand Index (RI) and Adjusted Rand Index (ARI)~\cite{sahlgren2005introduction}, as well as internal validation metrics including compactness (CP), the Dunn Validity Index (DVI) \cite{bezdek1998some}, silhouette score \cite{batool2021clustering}, and clustering stability measured via label alignment using the Hungarian algorithm \cite{liu2022stability}. Descriptions and formulas for each metric are provided in Section ~\ref{METHODS}.

To handle instability in Qlustering, we added an eighth step, invoked when runs yielded inconsistent decisions, to improve reliability: a consensus clustering protocol. This involved running the algorithm ten times on identical data and computing both the mean RI and ARI. In addition, we used hierarchical clustering average linkage to assign final groups from the consensus matrix following the technique Monti et. al. used in their work \cite{monti2003consensus} (see Fig.~\ref{fig1}b and Fig.~\ref{fig:OverlapBarrier}b). Further details on this protocol are provided in Sec.~\ref{METHODS}.

\section{RESULTS}
\label{RESULTS}
\subsection{Clustering of Points in 3D space}
\label{Position}

Our first example, though presented using state vectors, demonstrates a generalizable case applicable to diverse
data types. The inputs, $\Psi$, are vectors of length $d$, normalized such that their norm is unity, i.e. all points lie on the $(d-1)$-dimensional unit sphere. For demonstration, the $N$ input vectors are constructed as follows. We define $q$ ``base points'' $\{b_i\},\, i=1,\ldots,q$, and partition the $N$ data points into $q$ equal groups. In each group, vectors are drawn from a distribution centered on one of the base points with width parameter $0<\omega<1$. Each point is generated as $\Psi_n = (1-\omega)\,b_i + \omega\,u_n,$
where $u_n$ is a random unit vector, followed by normalization onto the unit sphere. Clearly, if $\omega$ is much smaller than the typical distance between base points, the groups remain well separated; for sufficiently large $\omega$, the groups overlap. At $\omega=1$, points are uniformly distributed on the unit sphere.

\begin{widetext}
    
\begin{figure}
    \centering
    \includegraphics[width=0.9\linewidth]{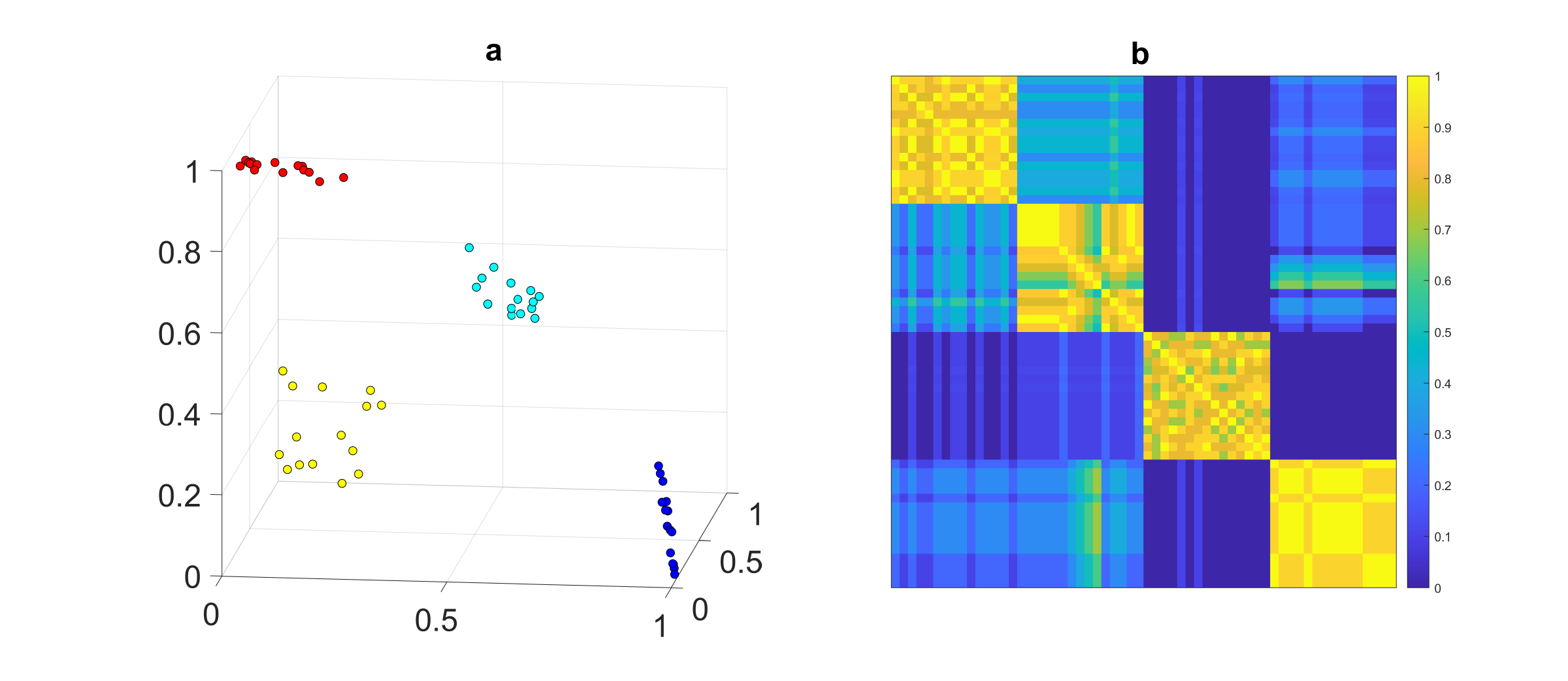}
\caption{\textbf{Qlustering of 3-dimensional vectors into four groups at $\omega = 0.15$.} 
(a) Spatial distribution of the input state vectors. (b) Consensus matrix from 10 repeated Qlustering runs. Yellow regions indicate high co-clustering frequency, while blue denotes low agreement. Each square represents a vector pair, with color intensity indicating the frequency of co-clustering (yellow: high consistency; blue: low agreement). A stable Qlustering pattern is observed, with four distinct block structures emerge in the consensus matrix, reflecting strong consistency in group assignments across runs.
Qlustering of 2-dimensional vectors into three groups at $\omega = 0.2$.} 
    \label{fig1}
\end{figure}
\end{widetext}

We evaluated Qlustering on 3-dimensional synthetic data. Overlap began at $\omega \simeq 0.3$. Perfect clustering was obtained for small $\omega$, decreasing to $\mathrm{RI} = 0.85$ and $\mathrm{ARI} = 0.63$ at $\omega = 0.3$ (Fig.~\ref{fig1}).

\begin{figure}[t]
    \centering
    \includegraphics[width=\linewidth]{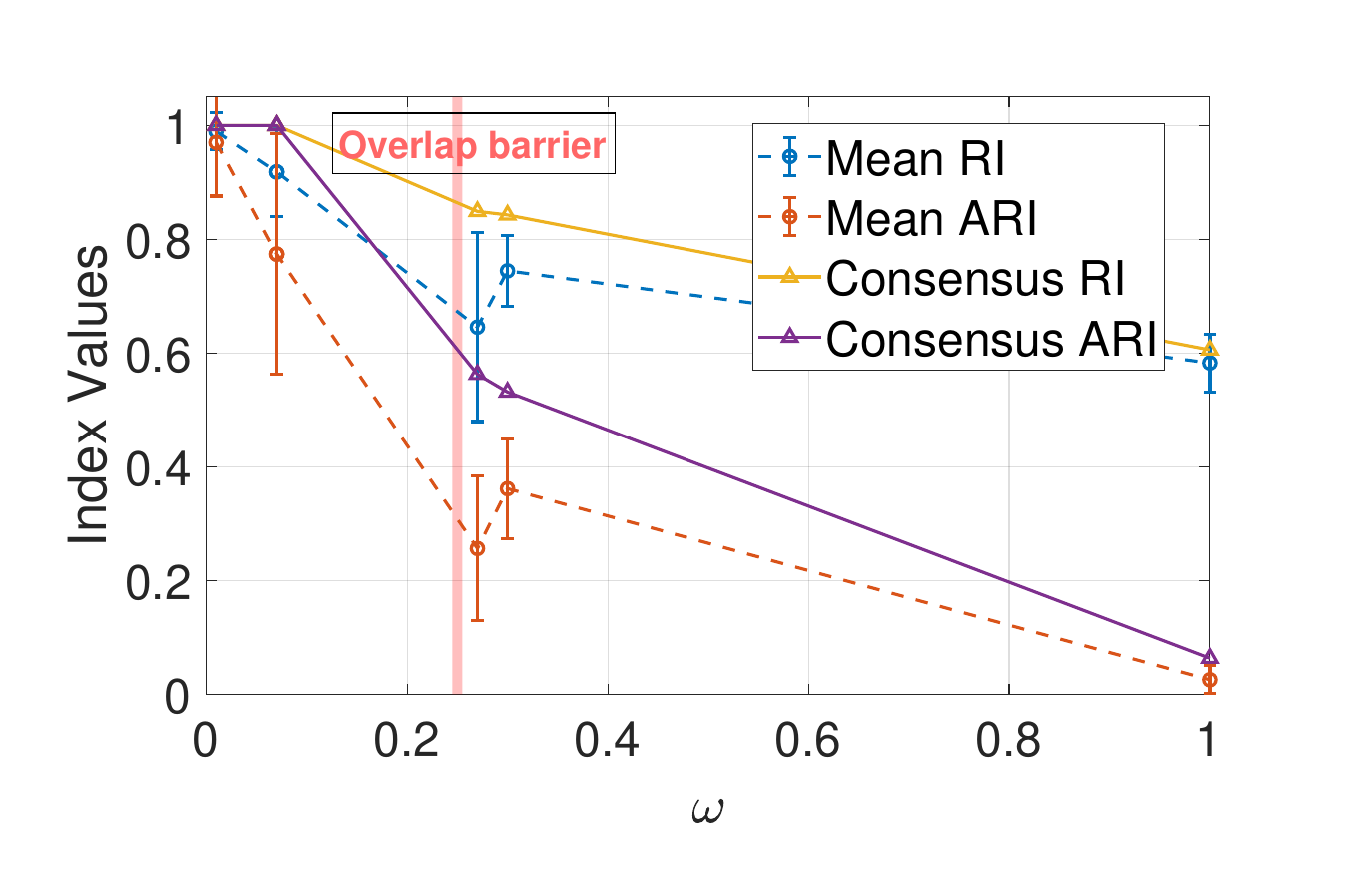}\\[0.5em] 
    \includegraphics[width=\linewidth]{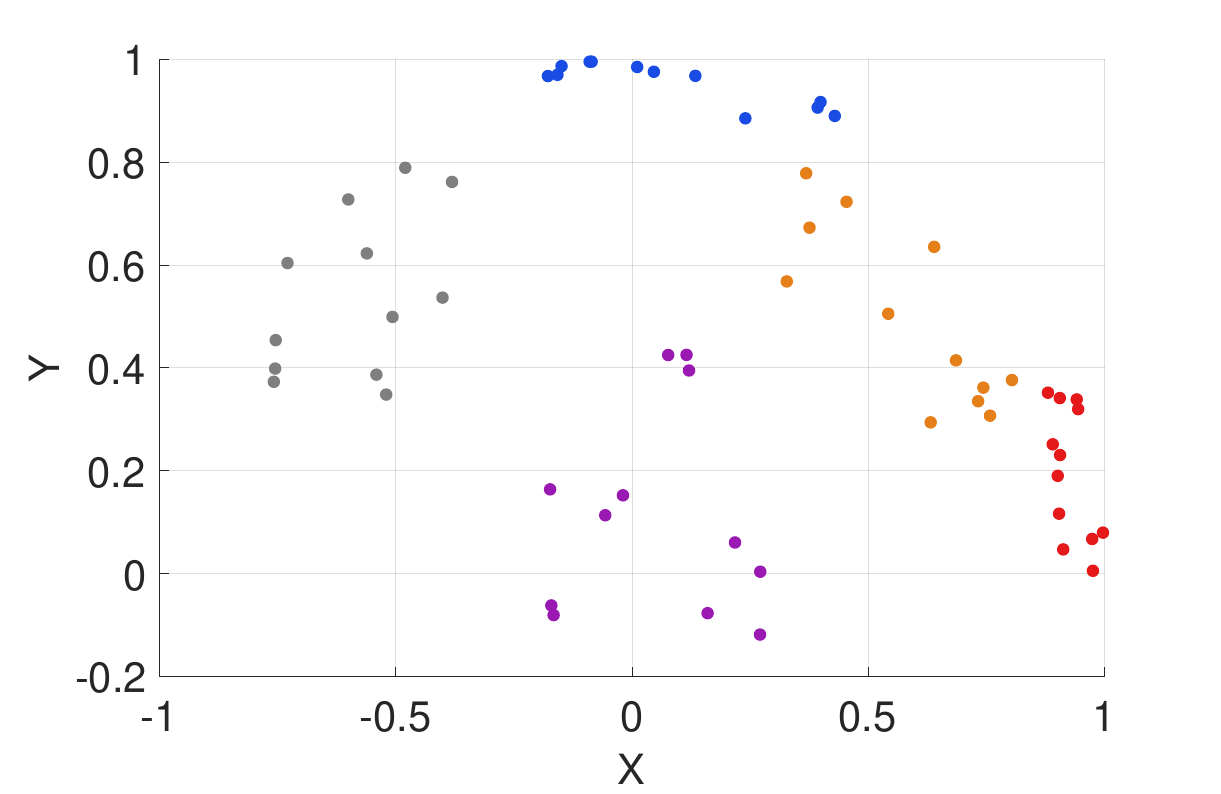}       
    \caption{\textbf{Top: Qlustering scores across varying group widths, $\omega$.} Solid-line triangles indicate the consensus clustering scores for the Rand Index (RI) and Adjusted Rand Index (ARI), while dashed-line circles represent the corresponding mean values. High clustering performance is maintained until significant overlap between groups occurs (marked by the transparent red line), after which scores decline toward random clustering levels as the distribution approaches uniformity. \textbf{Bottom: The overlap point visualized.} Spatial distribution of five groups in a three-dimensional Hilbert space at $\omega = 0.25$, marking the onset of the overlap point. Colors distinguish the clusters. The groups remain largely separated---i.e., no true overlap yet---though some points already lie at approximately equal distances from their own center and that of another group. This regime marks the transition where Qlustering performance begins to decline from perfect scores toward reduced accuracy, as shown in the top figure.}
    \label{fig:OverlapBarrier}
\end{figure}

To further challenge the algorithm and demonstrate Qlustering's capabilities with a larger number of classes in higher dimensions, we evaluated it on five groups ($q=5$) in a three-dimensional Hilbert space (Fig.~\ref{fig:OverlapBarrier}). Sixty data points were centered at $\nya{0,1,0}$, $\nya{0,0,1}$, $\nya{1,0,0}$, $\tfrac{2 \sqrt{3}}{9}\nya{-1.5,1.5,1.5}$, $\tfrac{2}{\sqrt{13}}\nya{0,1,1.5}$
, with varying group widths. For small widths, Qlustering achieved perfect clustering ($\mathrm{RI} = \mathrm{ARI} = 1$), while increasing the width reduced performance. Using the consensus scheme, perfect scores persisted over a wider range of widths, with performance declining only once substantial overlap occurred.

\subsection{Localization}
\label{Localization}

To demonstrate the versatility of \textbf{Qlustering} and its ability to perform on physical, high-dimensional data, we applied it to the localization problem described also in \cite{lorber2024using}. The dataset comprises two classes of quantum states, distinguished by their Inverse Participation Ratio (IPR) \cite{hikami1986localization}
\begin{equation}
    \mathcal{ IPR}[\psi]=\left(\sum^{L}_{i=1}|\psi(i)|^4\right)^{-1}
\end{equation} 
The IPR quantifies the spatial extent of a quantum state, with higher values indicating strong localization on a few sites and lower values corresponding to delocalized states spread over many sites.
To adapt Qlustering to this task, we modified the cost function as:  
\setlength{\abovedisplayskip}{5pt}
\setlength{\belowdisplayskip}{5pt}
\begin{equation*}
J_{\text{tags}} =
\begin{cases} 
[1; 0], & 0.4 > J_1 > 0.5 \\
[0; 1], & 0.4 > J_2 > 0.5 \\
[0.5; 0.5], & 0.4 \leq J_1, J_2 \leq 0.5
\end{cases}
\end{equation*}
\begin{equation}
\text{Cost} = \sum \|\mathbf{J_{\text{tags}}} - \mathbf{J} \|^2
\label{IPR cost}
\end{equation}

This formulation assigns extreme current values to one class and moderate (ambiguous) values to the other, effectively separating strongly localized states from delocalized ones. Each group spans a range of IPR values, leaving a gap between them defined as $\Delta_{\mathrm{IPR}} = IPR^{\mathrm{Highest}}_1 - IPR^{\mathrm{Lowest}}_2$, where $IPR^{\mathrm{Highest}}_1$ and $IPR^{\mathrm{Lowest}}_2$ denote the most extreme IPR values in each group. The algorithm attains high accuracy when $\Delta_{\mathrm{IPR}}$ is large and the classes are well separated. As the overlap between groups increases, $\Delta_{\mathrm{IPR}}$ diminishes, leading to a decline in classification accuracy.

\begin{figure}[t]
    \centering
    \includegraphics[width=\linewidth]{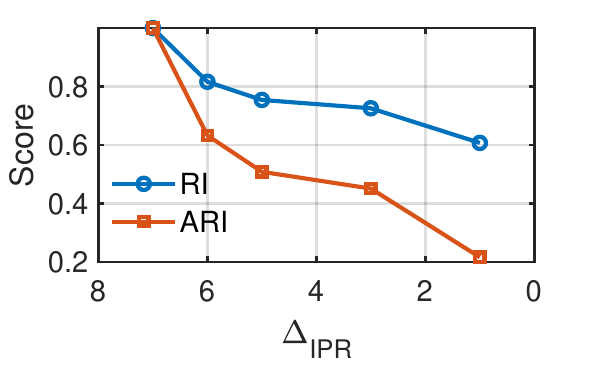}
    \caption{\textbf{Qlustering performance on the localization task.} Rand Index (RI) is shown in blue and Adjusted Rand Index (ARI) in orange on varying $\Delta_{IPR}$. The network clusters 10-dimensional wavefunctions with IPR values ranging from 1 to 10. As $\Delta_{IPR}$ decreases, performance declines. This drop corresponds to the mixing of strongly localized and delocalized states, which reduces the distinctiveness of the underlying physical classes.}

    \label{fig IPR}
\end{figure}

Figure~\ref{fig IPR} presents the RI and ARI as functions of class separability.

\subsection{QM9 Data Set}
\label{QM9}
To evaluate Qlustering on complex, high-dimensional chemical data, we used the QM9 benchmark dataset, which contains 133{,}885 small molecules with quantum‐chemical annotations \cite{ramakrishnan2014quantum,ruddigkeit2012enumeration}. Each entry includes a SMILES representation \cite{weininger1988smiles}, HOMO–LUMO energy gaps, temperature‐dependent internal energies, and other descriptors. Our objectives were: (i) to assess Qlustering’s robustness on realistic chemical informatics data, and (ii) to identify descriptor parameters that co-vary, potentially revealing structure–property relationships. Since Qlustering clusters points based on Hilbert‐space locality, the resulting groupings may reflect molecular features - such as symmetry or topological motifs - that influence specific chemical properties.

Molecules were encoded using their Sorted Interatomic Distances (SID), computed from the 3D coordinates $\mathbf{r}_i \in \mathbb{R}^3$ of all non-hydrogen atoms:
\[
\mathrm{SID} = \mathrm{sort}\Bigl(\bigl\{ \lVert \mathbf{r}_i - \mathbf{r}_j \rVert_2 
\;\big|\; i < j,\ Z_i, Z_j \neq 1 \bigr\}\Bigr),
\]
where $Z_i$ denotes the atomic number of atom $i$. This permutation‐invariant descriptor compactly encodes molecular geometry and has shown strong performance in small‐molecule regression tasks \cite{Hansen2015}. Each molecule was thus represented by a fingerprint vector of length~10, corresponding to the maximum number of non-hydrogen interatomic pairs observed among the 50 molecules selected from the dataset. Shorter vectors were zero-padded to this length to ensure uniform dimensionality.


The molecular fingerprint vectors were normalized into state vectors $\psi$ satisfying $\sum_j \psi(j)^2 = 1$. From the full dataset, a subset of 50 vectors was selected as input to Qlustering. Clustering was first performed with $q=2$ groups and, based on the consensus heatmap (Fig.~\ref{fig QM9 consensus}), also tested with $q=4$. For each case, 10 independent Qlustering runs were conducted, following the procedure in Sec.~\ref{Position}.

\begin{figure}[!t]
\centering
\includegraphics[width=\linewidth]{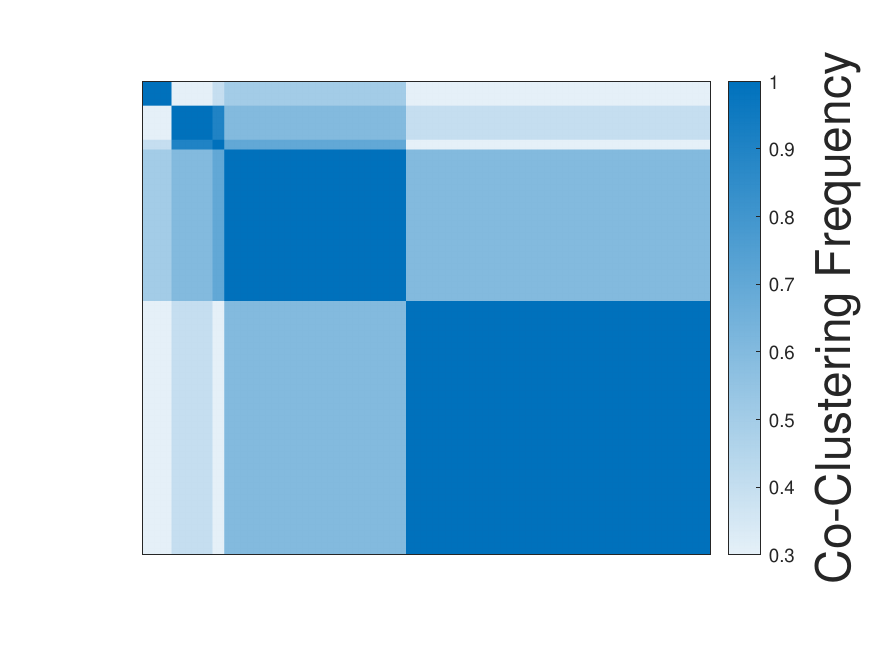}
\caption{\textbf{Consensus heatmap for Qlustering on a QM9 subset.} Pairwise consensus values from 10 consecutive Qlustering runs on a subset of 50 QM9 molecules are shown. Darker blue indicates a higher tendency for molecule pairs to cluster together, while lighter blue indicates lower co-clustering frequency. The network architecture (10--3--2) outputs two clusters; however, the consensus heatmap reveals a latent four-cluster structure. Running Qlustering with $q = 4$ confirmed this pattern and yielded improved internal clustering metrics, as discussed in the main text.}
\label{fig QM9 consensus}
\end{figure}

\begin{figure}[!t]
\centering
\includegraphics[width=1\linewidth]{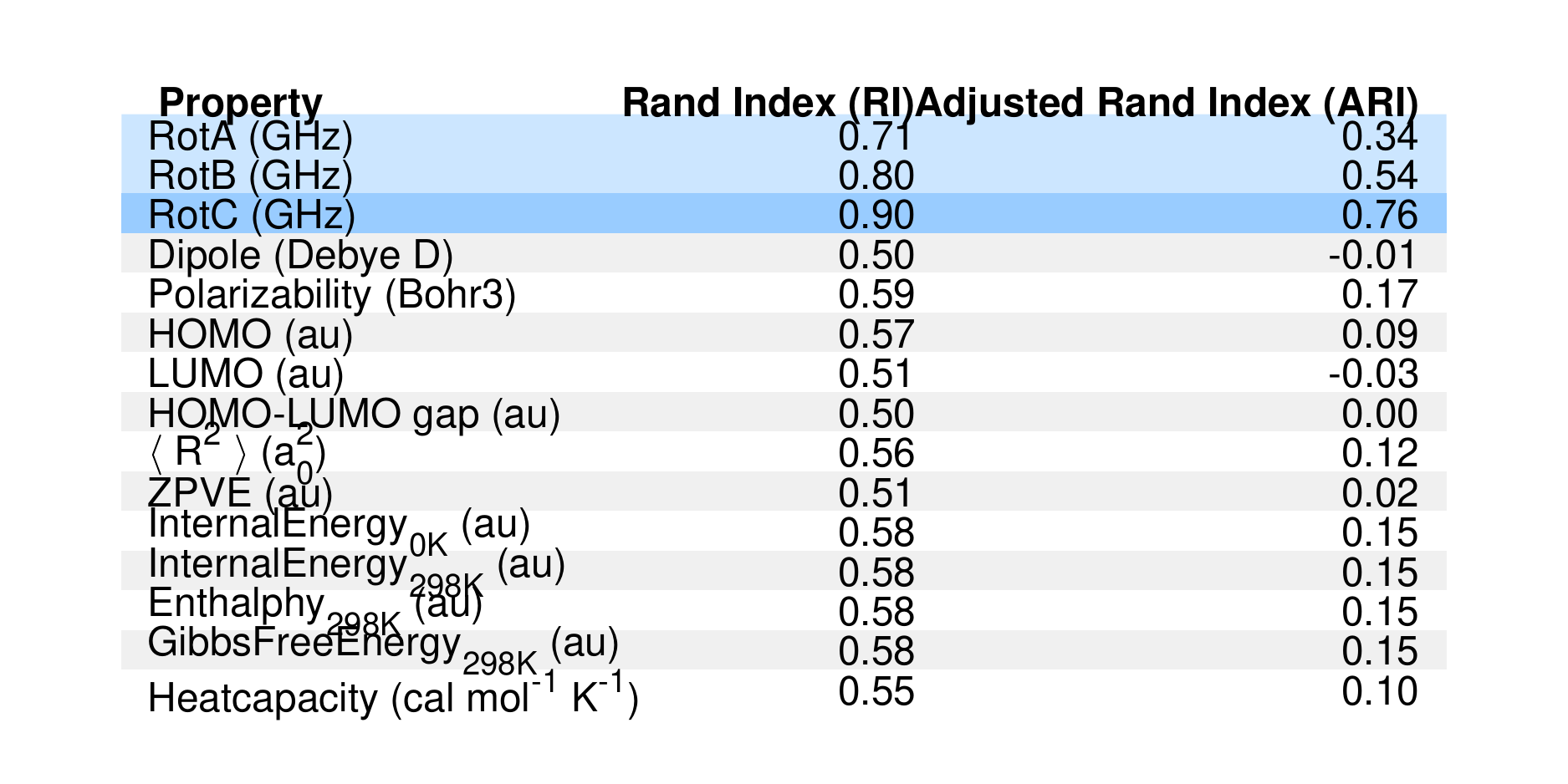}
\caption{\textbf{External clustering scores for 14 molecular properties from the QM9 dataset.}  
Best Rand Index (RI) and Adjusted Rand Index (ARI) values are shown for each binarized property, computed from Qlustering results on a subset of 50 molecules. Rotational constants A, B, and C yield the highest agreement with known labels, with RotC achieving $\mathrm{RI} = 0.90$ and $\mathrm{ARI} = 0.76$. These properties depend directly on molecular geometry, consistent with Qlustering’s structural input.}
\label{fig QM9 properties}
\end{figure}

Clustering results were evaluated using external metrics (RI, ARI) and internal metrics (compactness, Dunn Index, silhouette, stability), following Sec.~\ref{METHODS}. External scores for 14 binarized molecular parameters are shown in Table~\ref{fig QM9 properties}. The highest agreement was obtained for the rotational constant C (RotC), with best $\mathrm{RI} = 0.90$ and $\mathrm{ARI} = 0.76$, while rotational constants A and B also scored highly ($\mathrm{RI} > 0.70$, $\mathrm{ARI} > 0.30$). Rotational constants $(A, B, C)$, given by  
$A = \frac{h}{8\pi^2 c I_a}$,  
$B = \frac{h}{8\pi^2 c I_b}$,  
$C = \frac{h}{8\pi^2 c I_c}$,  
are inversely proportional to the principal moments of inertia $I_{a,b,c}$, where $h$ is Planck’s constant and $c$ is the speed of light. As these moments depend on the spatial distribution of atomic masses, the constants provide a direct probe of molecular geometry. This aligns with our aim of using Qlustering as a covariance detector, identifying correlations between structural features and physical properties.
Stability across runs reached $0.754$, exceeding typical benchmarks \cite{fahad2014survey}. Internal scores for $q = 2$ were: compactness $= 19.87$, Dunn Index $= 0.765$, silhouette $= 0.684$. Consensus clustering for RotC gave $\mathrm{RI} = 0.75$, $\mathrm{ARI} = 0.30$. Extending to $q = 4$ (no labels) yielded compactness $= 0.865$, Dunn Index $= 2.4251$, silhouette $= 0.9776$, and stability $= 0.80$ (see Fig.~\ref{fig QM9 consensus}).

\subsection{Computational complexity}
To assess the computational performance of the proposed Qlustering algorithm, we compare its complexity to that of classical clustering methods, particularly centroid-based approaches such as \textit{k}-means.

The \textit{k}-means algorithm exhibits a per-iteration complexity of \(\mathcal{O}(n \cdot k \cdot d)\), where \(n\) is the number of data points, \(k\) the number of clusters, \(d\) the dimensionality of the feature space, and \(i\) the number of iterations until convergence. The total complexity is therefore:
\[
\mathcal{O}(n \cdot k \cdot d \cdot i).
\]
While \textit{k}-means is efficient in low-dimensional, well-separated data, its performance degrades with increasing dimensionality and non-convex cluster shapes.

In contrast, Qlustering operates on a quantum-inspired framework. Each iteration consists of (i) propagating \(N\) quantum state vectors \(\Psi_n\), (ii) measuring their steady-state currents \(J[\Psi_n]\), and (iii) perturbing the system Hamiltonian \(\mathcal{H}\). The dominant computational cost is typically the measurement step, with per-iteration cost:
\[
\mathcal{O}(N \cdot T_m),
\]
where \(T_m\) is the time required for a single current measurement.

Furthermore, Qlustering updates \(\mathcal{H}\) by perturbing it using \(p\) particles per iteration, each contributing to the search in the Hamiltonian landscape. The time required per perturbation is denoted \(T_H\). Thus, the full per-iteration complexity becomes:
\[
\mathcal{O}(N \cdot T_m + p \cdot T_H),
\]
and total complexity over \(i\) iterations is:
\[
\mathcal{O}\left(i \cdot (N \cdot T_m + p \cdot T_H)\right).
\]

A key advantage of Qlustering lies in its rapid convergence. The cost function depends sensitively on the steady-state current \(J[\Psi_n]\), which closely reflects the eigenstructure of \(\mathcal{H}\). As a result, small perturbations in \(\mathcal{H}\) can lead to significant improvements in the cost function, typically resulting in a very small number of iterations (\(i \ll n\)).

In practice, the term \(p \cdot T_H\) may be minimized through two avenues: (i) the use of efficient hardware for Hamiltonian manipulation - such as Mach-Zehnder interferometers in photonic implementations - and (ii) problem-adaptive selection of \(p\), where fewer particles suffice to explore the search space. Even in less favorable scenarios, this term does not dominate, and the overall complexity remains tractable.


\subsection{Iris Data Set}
\label{iris}

The Iris dataset, introduced by Ronald A.~Fisher in 1936, is a classic benchmark in pattern recognition and unsupervised learning. It contains 150 samples from three iris species (\textit{Iris setosa}, \textit{Iris versicolor}, and \textit{Iris virginica}), each described by four features: sepal length, sepal width, petal length, and petal width. Although the dataset includes labeled classes, it is widely used in clustering tasks to assess an algorithm’s ability to recover natural groupings without prior knowledge. Its well-structured yet partially overlapping class distributions make it a useful testbed for comparing clustering methods.

\begin{figure*}
      \centering

  \subfloat{\includegraphics[width=0.48\textwidth]{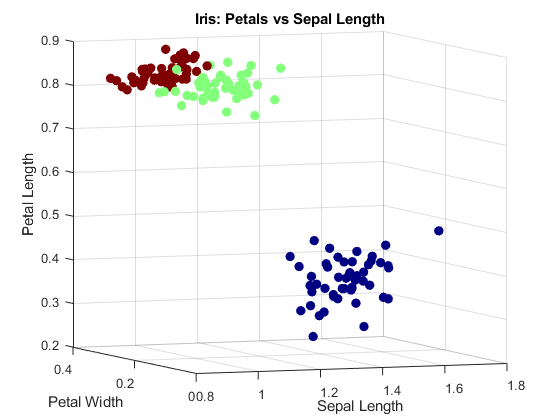}}
  \hfill
  \subfloat{\includegraphics[width=0.48\textwidth]{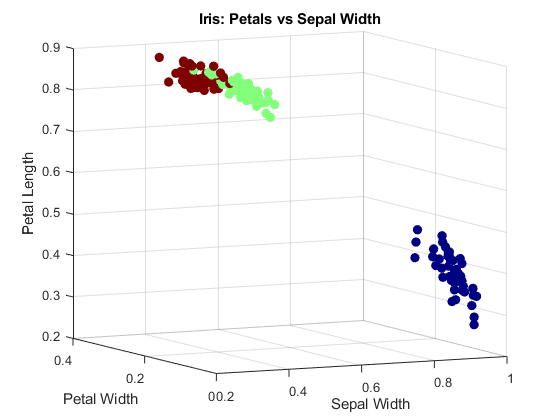}}

  \vspace{0.9em}

  \subfloat{\includegraphics[width=0.48\textwidth]{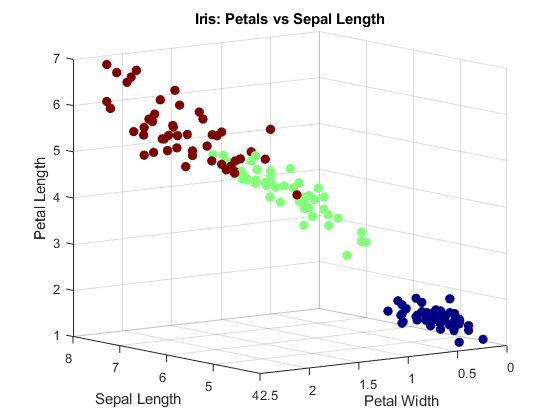}}
  \hfill
  \subfloat{\includegraphics[width=0.48\textwidth]{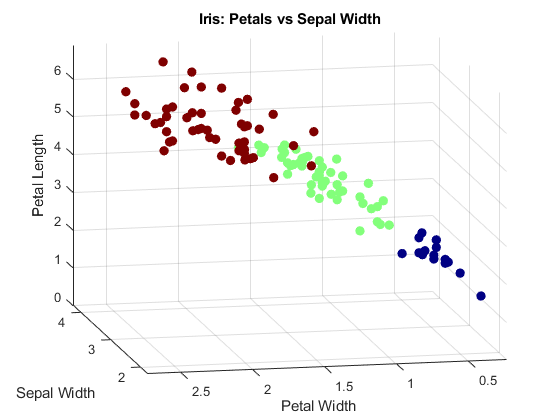}}

  \caption{\textbf{Feature distributions in the Iris dataset.}
  \textbf{(a)} Normalized, sepal length (less mixed);
  \textbf{(b)} Normalized, sepal width (mixed);
  \textbf{(c)} Unnormalized, sepal length (less mixed);
  \textbf{(d)} Unnormalized, sepal width (mixed).}
  \label{fig:iris}
\end{figure*}

The dataset was clustered using Qlustering, with internal and external evaluation scores measured over 10 consecutive runs. A neural network with a 4--2--3 architecture (four input neurons, two intermediate layers with one neuron each, and three output neurons) was used to learn feature representations conducive to clustering. Normalization transformed the input into state vectors, which skewed the clusters (Fig.~\ref{fig:iris}) and increased the difficulty of separation. To address this, the sepal width feature was removed - a common practice when clustering the Iris dataset \cite{mani2024enhancing} - yielding a substantial improvement in consensus clustering performance, outperforming k-means “on its own field.”

Using all four features, Qlustering achieved $\mathrm{RI} = 0.77$, $\mathrm{ARI} = 0.56$, and internal metrics of stability $= 0.72$, compactness (CP) $= 458.3$, Dunn Validity Index (DVI) $= 0.038$, and silhouette score $= 0.14$ (all mean values over 10 runs). Removing the sepal width feature increased external scores to $\mathrm{RI} = 0.92$ and $\mathrm{ARI} = 0.82$, with internal metrics of stability $= 0.60$, CP $= 2.5$, DVI $= 0.022$, and silhouette score $= 0.38$.

This issue highlights a potential weakness of Qlustering - the normalization step. However, with appropriate data preparation, this step can be mitigated and may even become an advantage for certain data structures.\\
\subsection{Assessment of k-means algorithm on the data}
\label{Classic}
Table~\ref{k-means} presents the internal and external evaluation metrics of the widely used k-means algorithm, applied to the same four clustering tasks explored with Qlustering in this study. As expected from a well-established and extensively tested method, k-means performs competitively in most cases. In fact, the two algorithms often yield comparable scores. However, several key differences in their behavior and strengths are worth noting:
\begin{itemize}[topsep=2pt, itemsep=2pt, parsep=0pt, leftmargin=10pt]
    \item \textbf{Compactness (CP):} K-means consistently outperforms Qlustering in terms of compactness. This is expected, as CP serves as the cost function optimized by k-means itself, giving it a natural advantage on this metric.
    
    \item \textbf{Position problem:} In this synthetic setup, k-means achieves strong results even when $\omega = 0.3$, where the group boundaries begin to overlap. This suggests that Qlustering may struggle with convex or linearly separable data structures under certain configurations.
    
    \item \textbf{Localization problem:} In contrast, k-means performs poorly - almost randomly - on the localization task. This is consistent with known limitations of k-means in high-dimensional data spaces \cite{fahad2014survey}. Qlustering, in comparison, only shows degraded performance when the inter-group parameter range (IPR) gap is minimal.
    
    \item \textbf{QM9 dataset:} On this real-world dataset, k-means shows strong internal metrics but weak external agreement with known labels. This discrepancy further highlights the strength of Qlustering in uncovering meaningful parameter dependencies beyond geometric compactness.
    
    \item \textbf{Iris dataset:} With all four features included, k-means outperforms \textbf{Qlustering} on most metrics. However, removing the sepal width feature - known to contribute to class overlap - reverses the outcome: Qlustering surpasses k-means on the majority of external and internal scores.

\end{itemize}
\begin{figure}[!t]
\centering
\includegraphics[width=\linewidth]{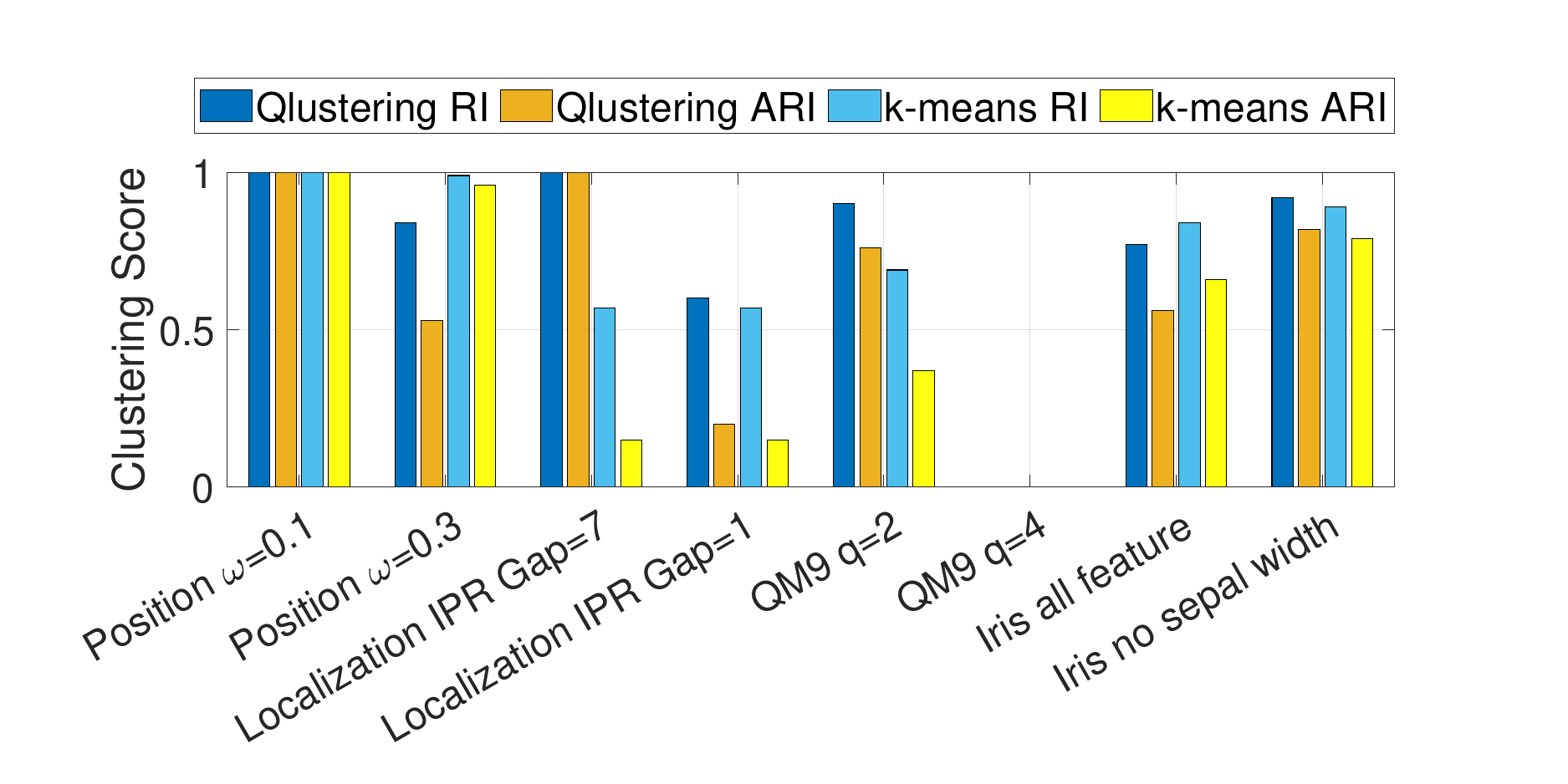}
\caption{\textbf{External clustering performance of Qlustering vs.\ k-means.}  
Rand Index (RI) and Adjusted Rand Index (ARI) scores are shown for both algorithms across multiple datasets: positional clustering with $\omega = 0.1$ and $\omega = 0.3$, localization with IPR gaps of 7 and 1, QM9 with $q = 2$ and $q = 4$, and the Iris dataset with all features and with sepal width removed}

\label{fig QM9 consensus}
\end{figure}
\onecolumngrid 

\begin{table*}[h] 
    \centering
    \caption{K-Means Clustering Performance Scores}
    \label{tab:kmeans_scores}
    \begin{tabular}{|l|c|c|c|c|c|c|c|}
        \hline
        \textbf{Score} & \multicolumn{2}{c|}{\textbf{Position}} & \multicolumn{2}{c|}{\textbf{Localization}} & \multicolumn{2}{c|}{\textbf{QM9}} & \textbf{Iris} \\
        \cline{2-8}
        & \textbf{$\omega=0.1$} & \textbf{$\omega=0.3$} & \textbf{IPR Gap$=7$} & \textbf{IPR Gap$=1$} & \textbf{$q=2$} & \textbf{$q=4$} &  \\
        \hline
        \multicolumn{8}{|l|}{\textbf{External Scores}} \\
        \hline
RI & $1$ & $0.987$ & $0.571$ & $0.571$ & $0.693$ & $\,\_\_\_$ & $0.843$ \\
ARI & $1$ & $0.957$ & $0.150$ & $0.150$ & $0.373$ & $\,\_\_\_$ & $0.659$ \\
\hline
\multicolumn{8}{|l|}{\textbf{Internal Scores}} \\
\hline
CP & $0.482$ & $8.31$ & $14.109$ & $9.73$ & $15.434$ & $1.198$ & $91.905$ \\
DVI & $3.337$ & $0.132$ & $0.415$ & $0.257$ & $0.674$ & $2.188$ & $0.095$ \\
Silhouette & $0.981$ & $0.630$ & $0.417$ & $0.345$ & $0.697$ & $0.716$ & $0.716$ \\
\hline
Stability & $1.00$ & $0.790$ & $0.846$ & $0.750$ & $0.851$ & $0.940$ & $0.851$ \\

        \hline
    \end{tabular}
    \label{k-means}
\end{table*}
\twocolumngrid

\section{METHODS}
\label{METHODS}
\textbf{Current from the Lindblad Equation}

We consider a system governed by a general tight-binding Hamiltonian, with dynamics described by the Lindblad master equation:
\begin{eqnarray}
\dot{\rho} = -i[\cH, \rho] + \sum_k \left( V_k^{\dagger} \rho V_k - \frac{1}{2} { V_k^{\dagger} V_k, \rho } \right) \nonumber \
= -i[\cH, \rho] + \mathcal{L}[\rho]~. \nonumber
\end{eqnarray}
The calculation is restricted to the single-exciton manifold, which suffices for capturing the relevant physics, as established in prior work \cite{zerah2021photosynthetic}.

We solve for the steady-state density matrix $\rho_s$, defined by $\dot{\rho}_s = 0$. The current $J_i$ at site $i$ is given by the continuity equation,
\begin{eqnarray}
J_i = \frac{d\langle n_i \rangle}{dt} = \frac{d}{dt} \mathrm{Tr}(\hat{n}_i \rho)~.
\end{eqnarray}
while $d\langle ... \rangle$ means average value over the network sites and $n_i$ is the number density in site $i$.
Taking the time derivative inside the trace yields:
\begin{eqnarray}
J_i = \mathrm{Tr}(\dot{\hat{n}}_i \rho_s + \hat{n}_i \dot{\rho}_s)~.
\label{eq:Ji}
\end{eqnarray}

At steady state the expression simplifies. For exit sites, where $\dot{\hat{n}}_i = -i[\hat{n}_i, \cH]$, the total current conservation implies:
\begin{eqnarray}
J_{ext} = \mathrm{Tr} \left( \hat{n}_{ext} \left( -i[\cH, \rho_s] + \mathcal{L}_{ext}[\rho_s] \right) \right) = 0~.
\label{eq:Jcons}
\end{eqnarray}

Due to current conservation, at steady state the current outside the network equals in magnitude to the current inside, and opposite in sign. Hence the zero net current. The outside current $J_{ext}$ can be evaluated solely from the term:
\begin{eqnarray}
J_{ext} = \mathrm{Tr} \left( \hat{n}_{ext} \mathcal{L}_{ext}[\rho_s] \right)~.
\label{eq:Jfinal}
\end{eqnarray}
\\
\textbf{External and Internal Validation Metrics}
In this subsection we will shortly provide the formulas and key-features of the validation metrics used in this paper. for deeper reading, please see the include papers cited near each method\\
\paragraph{Random Index (RI) and Adjusted Random Index (ARI) scores} The RI measures the proportion of pairwise agreements,whether two points are correctly grouped or separated, and is defined as:
\[
\text{RI} = \frac{TP + TN}{TP + TN + FP + FN}
\]
where \( TP \) and \( TN \) denote true positives and true negatives, and \( FP \), \( FN \) are false positives and false negatives. RI values range from 0 (complete disagreement) to 1 (perfect agreement).

However, RI does not account for agreement due to chance; high RI values may still occur in poorly performing models. To address this, we use the ARI:
\[
\text{ARI} = \frac{\text{RI} - \mathbb{E}[\text{RI}]}{\max(\text{RI}) - \mathbb{E}[\text{RI}]}
\]
where \( \mathbb{E}[\text{RI}] \) is the expected RI of a random model. ARI ranges from \(-1\) to 1, with 0 indicating agreement by chance and 1 for a perfect match. Negative values mean the clustering is worse than random.
\paragraph{Compactness (CP)}  
Compactness measures the within-cluster dispersion by computing the sum of squared distances between data points and their respective cluster centroids:
\[
\text{CP} = \sum_{i=1}^{N} \left\| \mathbf{x}_i - \mathbf{c}_{\text{label}(i)} \right\|^2
\]
where \( \mathbf{c}_{\text{label}(i)} \) denotes the centroid of the cluster to which point \( \mathbf{x}_i \) is assigned. Lower CP values indicate tighter, more cohesive clusters.

\paragraph{Dunn Validity Index (DVI)}  
The Dunn Index evaluates the ratio between the minimum inter-cluster distance and the maximum intra-cluster diameter:
\[
\text{DVI} = \frac{ \displaystyle \min_{i \neq j} \delta(C_i, C_j) }{ \displaystyle \max_{k} \Delta(C_k) }
\]
where:
\begin{itemize}
    \item \( \delta(C_i, C_j) \) is the minimum pairwise distance between points in clusters \( C_i \) and \( C_j \),
    \item \( \Delta(C_k) \) is the maximum distance between any two points within cluster \( C_k \).
\end{itemize}
Higher DVI values indicate well-separated and compact clusters.

\paragraph{Silhouette Score}  
The silhouette score captures how similar an object is to its own cluster compared to other clusters. For each point \( \mathbf{x}_i \), the silhouette value \( s(i) \) is defined as:
\[
s(i) = \frac{b(i) - a(i)}{\max\{a(i), b(i)\}}
\]
where:
\begin{itemize}
    \item \( a(i) \) is the average intra-cluster distance (cohesion),
    \item \( b(i) \) is the lowest average inter-cluster distance to any other cluster (separation).
\end{itemize}
The overall silhouette score is the mean of all \( s(i) \). Values close to 1 indicate good clustering; values near 0 suggest overlapping clusters.

\paragraph{Stability via Label Alignment}  
To assess the stability of clustering across multiple runs, we align cluster labels using the Hungarian matching algorithm applied to pairwise confusion matrices. The stability metric is defined as the average alignment accuracy over all pairwise label permutations:
\[
\text{Stability} = \frac{2}{R(R - 1)} \sum_{i < j} \text{Match}(R_i, R_j)
\]
where \( R \) is the number of clustering repetitions, and \( \text{Match}(R_i, R_j) \) is the fraction of matching labels between runs \( i \) and \( j \) after optimal alignment.\\

\textbf{Consensus Clustering Procedure}

To ensure robust and stable group identification in unlabeled data, we employed \textbf{consensus clustering} following the protocol of Monti et al.~\cite{monti2003consensus}.

\textbf{Step 1: Repeated Clustering.} We ran the clustering algorithm \( R = 10 \) times on the same dataset using random initializations. Each run produced a partition \( P_r \), where \( r \in \{1, 2, ..., R\} \).

\textbf{Step 2: Consensus Matrix Construction.} An \( n \times n \) consensus matrix \( C \) was constructed as:

\[
C_{ij} = \frac{1}{R} \sum_{r=1}^{R} \mathbb{I}\left( x_i \text{ and } x_j \text{ belong to the same cluster in } P_r \right)
\]

where \( \mathbb{I} \) is the indicator function, and \( n \) is the number of data points.

\textbf{Step 3: Final Clustering.} We computed a distance matrix \( D = 1 - C \) and applied \textit{hierarchical clustering with average linkage} (UPGMA) to assign the final cluster labels.

\textbf{Rationale.} Averaging performance metrics (e.g., RI, ARI) over multiple runs does not reflect clustering stability. In contrast, consensus clustering encodes the frequency with which samples co-occur in the same cluster across runs, yielding more reliable, initialization-independent groupings and allowing for visual and quantitative stability assessment.

\section{SUMMERY AND DISCUSSION}
In this work, we introduced Qlustering, a novel unsupervised machine learning algorithm inspired by physical systems. We evaluated its performance on four distinct tasks: synthetic state vectors clustered in Hilbert space, localization-based clustering using the inverse participation ratio (IPR), molecular data from the QM9 chemical database, and the classical Iris dataset.

Qlustering was assessed using both internal and external clustering metrics, as well as a theoretical analysis of its computational complexity. For comparison, the classical k-means algorithm was also benchmarked (see Section~\ref{Classic}).
Using these diverse techniques, we investigated the capabilities of Qlustering under different circumstances:

\begin{itemize}
    \item \textbf{Computational complexity---} When the Qlustering algorithm is executed on a physical computational component, i.e., a quantum network modeled by the Lindblad equation, its run time is expected to be on the same order as that of $k$-means—hence, very fast. Assuming a smooth energy landscape that promotes rapid convergence, Qlustering may even outperform $k$-means by several orders of magnitude. The details of such possible implementation will be discussed shortly

    \item \textbf{Dataset size---} Dataset scaling is a crucial factor in current computational paradigms. In this work, we employed relatively small datasets (up to 150 data points) due to the long runtime of Qlustering on classical hardware. As the algorithm is inherently physical in nature, larger datasets have not yet been systematically tested.

    \item \textbf{High dimensionality---} In Sec.~\ref{Localization}, we applied Qlustering to IPR data with ten parameters per point. Despite this relatively high dimensionality, Qlustering exhibited strong performance,  exceeding that of classical $k$-means.

    \item \textbf{Physical robustness---} In contrast to existing quantum machine-learning methods, Qlustering is highly robust because it requires neither qubits nor quantum gates. This feature facilitates scalability with respect to both network size and dataset size—critical parameters in practical implementations.

    \item \textbf{Type of datasets---} The main limitation of Qlustering lies in its sensitivity to noisy or mixed-type data. Across all tested scenarios, classification performance decreased as the overlap between groups increased, in comparison with $k$-means. This suggests that Qlustering performs optimally on well-separated data.

    \item \textbf{Internal validity---} Internal validity depends strongly on the type of input data. See Sec.~\ref{Classic} for a detailed discussion.

    \item \textbf{External validity---} External validity reflects the overall consistency of the method. Qlustering demonstrates relatively high performance across most tasks, which improves further when combined with consensus clustering post-processing. Its advantage over classical algorithms such as $k$-means arises from its robustness to initialization: while $k$-means may converge to incorrect minima due to poor initialization, Qlustering rarely exhibits such behavior. This stems from its simple energy landscape, which directs most initial configurations toward the same or similar energy wells.
\end{itemize}

This study demonstrates the feasibility of Qlustering as a reliable, general-purpose approach for clustering and analyzing parameter dependencies. Future research may focus on refining cluster-number estimation, exploring diverse data types and preprocessing strategies, and optimizing network-specific hyperparameters to further enhance performance. With its robustness to noise—requiring neither qubits nor quantum gates—low computational complexity, and fast convergence, Qlustering is particularly well suited for clustering complex yet well-separated, high-dimensional data.

Regarding possible implementations, it is worth noting the work of Caruso \textit{et al.}~\cite{caruso2016fast}, who employed a photonic maze to realize random quantum walks, a concept closely related to the motion equations used here. Similarly, programmable photonic processors~\cite{harris2018linear,harris2017quantum} offer a promising platform for implementing Qlustering. Qlustering can also be realized on a general-purpose quantum computer, offering a gate-less computation through state-preparation, an avenue which is currently under pursue.     


\label{summary}
\appendix
\section{Clustering of points in 2D space}
\label{app. a}

In the two-dimensional case ($d=2$) with $N=60$ and three clusters ($q=3$) at $b_1= \nya{1,0}$, $b_2=\nya{\tfrac{1}{\sqrt{2}},\tfrac{1}{\sqrt{2}}}$,$b_3=\nya{0,1}$,
overlap began around $\omega\simeq 0.25$.

For $\omega = 0.05$, perfect clustering ($\mathrm{RI} = \mathrm{ARI} = 1$) was achieved in all 10 runs; at $\omega = 0.1$, mean scores dropped to $\bar{\mathrm{RI}} = 0.87$ and $\bar{\mathrm{ARI}} = 0.86$. Using the consensus scheme (Sec.~\ref{METHODS}), perfect scores persisted up to $\omega \leq 0.2$, with performance declining to $\mathrm{RI} = 0.84$, $\mathrm{ARI} = 0.65$ beyond this threshold.
\begin{widetext}
    
\begin{figure}
    \centering
    \includegraphics[width=\linewidth]{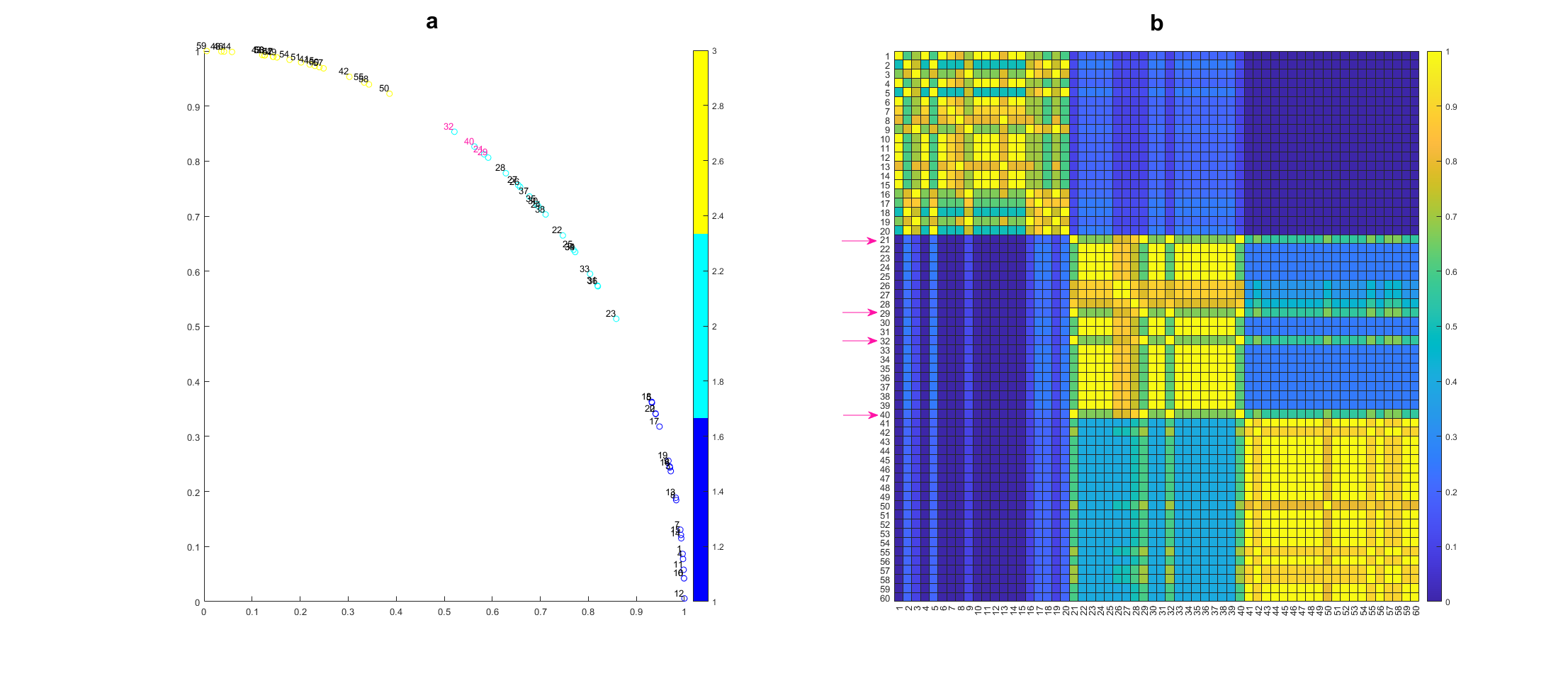}
\caption{\textbf{Qlustering of 2-dimensional vectors into three groups at $\omega = 0.2$.} 
(a) Spatial distribution of the input vectors. (b) Consensus matrix heatmap from 10 repeated Qlustering runs. Each square represents a vector pair, with color intensity indicating the frequency of co-clustering (yellow: high consistency; blue: low agreement). A stable Qlustering pattern is evident, as three distinct triangular blocks emerge in the consensus matrix. Pink arrows in (b), corresponding to the numbered markers in (a), indicate vectors frequently misclassified across runs, typically located near cluster boundaries. The quantum network employs a 2--3--3 node architecture.}
    \label{fig k=2}
\end{figure}

\end{widetext}
\textbf{\large Code availability}\\
The custom code that was created during the work that led to the main results of this article is published in a public GitHub repository: https://github.com/ShmueLorber/Qlustering 
\bibliography{refs}  
\end{document}